\documentstyle[epsfig]{mn}

\newcommand{\be}{\begin{equation}}
\newcommand{\ee}{\end{equation}}

\newcommand{\ls}{\raisebox{-.8ex}{$\buildrel{\textstyle<}\over\sim$}}

 \newcommand{\apj}{{\it ApJ, }}
 \newcommand{\apjl}{{\it ApJ(Letters), }}
 
 \newcommand{\aj}{{\it AJ, }}
 \newcommand{\mnr}{{\it MNRAS, }}

\newcounter{pp3}
\addtocounter{pp3}{3}

\title[Bending Waves in Accretion Discs]{Hydrodynamic Simulations of Propagating Warps and Bending Waves in Accretion Discs}

\author[Richard P. Nelson and John C.B. Papaloizou]{Richard P. Nelson 
\thanks{Email address:
R.P.Nelson@qmw.ac.uk} \& John C.B. Papaloizou \thanks {Email address:
J.C.B.Papaloizou@qmw.ac.uk} \\
 Astronomy Unit, 
 Queen Mary \& Westfield College, Mile End
 Rd, London E1 4NS}

 \date{Received *****; in original form ******}
 \pagerange{\pageref{firstpage}--\pageref{lastpage}}

 \def\LaTeX{L\kern-.36em\raise.3ex\hbox{a}\kern-.15em
	 T\kern-.1667em\lower.7ex\hbox{E}\kern-.125emX}

\begin{document}
 \label{firstpage}

 \maketitle

 \begin{abstract}
We present the results of a study of propagating warp or bending waves 
in accretion discs. Three dimensional hydrodynamic simulations were performed
using SPH, and the results are compared with calculations based on the linear
theory of warped discs.

\noindent We examine the response of a gaseous disc to an initially imposed
warping disturbance under a variety of physical conditions.
We consider primarily the physical regime in which the dimensionless
viscosity parameter $\alpha < H/r$, where $H/r$ is the disc aspect ratio,
so that bending waves are expected to propagate. We also performed
calculations for disc models in which $\alpha>H/r$, where the warps are
expected to evolve diffusively.  Small
amplitude (linear) perturbations are studied in both Keplerian and slightly non 
Keplerian discs, and we find that the results of the
SPH calculations can be reasonably
well fitted by  those of 
the linear theory. The main results of these
calculations are: ({\it i}) the warp in Keplerian discs 
when $\alpha < H/r$ propagates with
little dispersion, and damps at a rate expected from estimates of the
code viscosity, ({\it ii}) warps evolve diffusively when $\alpha > H/r$,
({\it iii}) the slightly non Keplerian discs lead to a substantially
more dispersive behaviour of the warps, which damp at a similar rate to the 
Keplerian case, when $\alpha < H/r$.

\noindent Initially imposed higher amplitude, nonlinear warping disturbances
were studied in Keplerian discs. The results indicate that nonlinear warps
can lead to the formation of shocks, and that the evolution
of the warp becomes less wave--like and more diffusive in character.

\noindent This work is relevant to the study of the
warped accretion discs that may occur around Kerr black holes or in misaligned
binary systems, and is mainly concerned with discs in which $\alpha < H/r$.
The results indicate
that SPH can model the hydrodynamics of
warped discs, even when using
rather modest numbers of particles.

 \end{abstract}

 \begin{keywords} accretion discs-bending waves-warps-SPH-simulations
 \end{keywords}

\section{Introduction}

Accretion discs occur in a variety of astrophysical contexts such as around 
T--Tauri stars, around Galactic X--ray sources such as Cygnus X--1, or
around massive black holes at the centres of active galactic nuclei.
In a situation where these discs are subject to out--of--plane
forcing, the disc may become globally twisted or warped. Examples
of where this may occur are in a binary
system where the binary orbit plane is misaligned with
respect to the disc midplane (e.g. Papaloizou \& Terquem 1995, Larwood {\em
et al.} 1996), or around a rotating black hole where the Lense--Thirring effect
may cause a warped disc to form (Bardeen \& Petterson 1975, Kumar \& Pringle
1985, Nelson \& Papaloizou 1999).

The linear theory of warped discs, in which both viscous and pressure effects
were included, was studied by Papaloizou \& Pringle (1983). They examined
the evolution of discs in the regime where $\alpha > H/r$,  
$\alpha$ being the dimensionless viscosity coefficient of Shakura \& Sunyaev 
(1973) and $H/r$ being the disc aspect ratio,  and showed that warps
evolve diffusively in this regime (see also Ogilvie 1999).
 The diffusion coefficient associated
with the warp evolution was found to be larger than that associated
with mass flow through the disc by a factor of $\sim 1/(2 \alpha^2),$
assuming an isotropic viscosity, indicating that disc warps diffuse more 
rapidly than had previously been supposed (e.g. Bardeen \& Petterson 1975).
For discs in which  $\alpha < H/r$, the governing equation for disc warp
evolution changes from being a diffusion equation
 to a wave equation, indicating that 
warps propagate as bending waves in this physical regime.  These bending waves
were studied using linear theory 
by Papaloizou \& Lin (1994, 1995), who showed under
 the assumption of very low frequency waves that
warping disturbances in a Keplerian disc
 propagate with no dispersion at a speed  corresponding to
an appropriate average of the 
sound speed.

The dynamics of warped discs have recently been studied using numerical
simulations performed with SPH. Larwood {\em et al.} (1996) examined the structure 
of a disc around one component of a binary system in which the disc midplane
and the binary orbit plane were misaligned. A similar study of warped
circumbinary discs was undertaken by Larwood \& Papaloizou (1997).
In a companion paper to this one (Nelson \& Papaloizou 1999), 
we study the dynamics
of warped accretion discs orbiting around rotating black holes, where
the angular momentum vector of the black hole is misaligned with that of
the disc.  In this case, the combined effects of differential Lense--Thirring 
precession due to the Kerr black hole and viscosity cause the 
inner regions
of these discs to become warped.

In this paper, we examine the evolution of warping disturbances in accretion 
discs using SPH in greater detail, and compare the results of the nonlinear
simulations with solutions to the linearized initial value
 problem obtained using a standard finite
difference method. The wavelengths of the warping perturbations that we study
are on the order of the disc thickness, so we relax the assumption of low 
frequency bending waves in the linear theory. We examine the evolution of warps
under a variety of physical conditions, including discs orbiting in both 
Keplerian and slightly non Keplerian central potentials, and warps with both 
small and large amplitudes. We find good agreement between the SPH simulations
and the predictions of linear theory for low amplitude warps.

In the regime where $\alpha < H/r$, the warps propagate across the disc as
inward and outward moving bending waves.
When $\alpha > H/r$, the warp evolution becomes diffusive, as expected.
We find that
the evolution of large amplitude warps becomes increasingly diffusive
and less wave--like
due to nonlinear effects such as shocks, which form because perturbed 
velocities on the order of the sound speed are generated.

The plan of this paper is as follows. In section (\ref{basic-eq}) we present
the basic equations of the problem. In section (\ref{linear})
we present a discussion of the linear theory of gaseous warped discs. 
In section (\ref{Num}) we describe the numerical method used to perform the
nonlinear simulations , and in section (\ref{waves}) we present the results of
these simulations and those obtained  from appropriate
 solutions  of the linearized
initial value problem. 
Finally, a discussion of our results and
our conclusions are  presented in section (\ref{concl}).

\section{Equations of motion} \label{basic-eq}
 In order to describe a compressible fluid, we adopt 
 the continuity and momentum equations  in the form
\be \frac{d \rho}{dt} + \rho \nabla.{\bf v} = 0 \label{cont}\ee
and
\be \frac{d{\bf v}}{dt} = - \frac{1}{\rho} \nabla P 
 - \nabla \Phi + {\bf S}_{visc}
 \label{moment}\ee
where 
\be \frac{d}{dt} = \frac{\partial}{\partial t} + {\bf v}.\nabla \ee
denotes the convective derivative, $\rho$ is the density, ${\bf v}$ is the 
velocity, $P$ is the pressure,  $\Phi$ is the gravitational potential
and ${\bf S}_{visc}$ represents the viscous 
force per unit mass. 

To describe the disc, we  adopt a cylindrical
coordinate system $(r, \phi, z)$ based on the central mass around which
the accretion disc orbits.  The corresponding Cartesian
coordinates are $(x,y,z).$ The distance to the central mass is given
by $R = \sqrt{r^2 + z^2}$, and the  position vector measured from there is
denoted by ${\bf r}$.

The gravitational potential $\Phi$ used in the calculations is given by
\be \Phi = -\frac{G M}{\sqrt{R^2 + b^2}} \label{phi-grav} \ee
where $b$ is an optional softening parameter used to prevent numerical
divergences in the gravitational force for accretion disc models that
extend to the centre of the coordinate system.

The equation of state is taken to be that of a polytrope with $\gamma=5/3$:
\be P = K \rho^{\gamma}. \label{estate} \ee
Energy dissipated through the action of artificial viscosity is 
simply allowed to leave the system, so that a barotropic equation of state
is assumed throughout.

\section{Linear Theory of Warps and Bending waves} \label{linear}
The linear theory of  warps 
in a  vertically stratified gaseous
accretion disc, including
hydrodynamical effects, was initially 
investigated by Papaloizou \& Pringle (1983).

The analysis presented in their paper, however, applied only to 
viscous accretion discs
in which the  dimensionless
Shakura \& Sunyaev (1973) viscosity parameter, $\alpha$, was constrained
to lie in the range $H/r < \alpha < 1$, where $H/r$ is the ratio of disc height 
to radius. 
Under these circumstances,  globally 
warped discs were found to evolve diffusively
on a time scale $t_D \sim \alpha^2 t_{\nu}$, where $t_{\nu} = r^2/{\nu}$ is the
viscous time scale appropriate to accretion,
 expected from standard thin disc theory.
We comment here that in the above a standard isotropic viscous
stress tensor was assumed.

For the regime $\alpha < H/r$, however, 
the nature of the governing equation for
the disc tilt changes from being of the diffusion type to being 
a wave equation,
such that warps  are communicated through  the disc {\em via} the
propagation of bending waves (Papaloizou \& Lin 1994, 1995).

\subsection{Initial Value Problem}
Small amplitude warps may be considered to be linear perturbations
of a disc with midplane  initially  coincident with
the $(x,y)$ plane.  The 
 $\phi$  dependence  of all perturbations
 may be taken into account through a factor
$e^{i(m \phi )}$, where 
$m$ is the azimuthal mode number. For the  global warps
considered here, $m=1$.
The components of the 
linearized equations of motion  for a barotropic
fluid   (see eg. Papaloizou \& Lin
1994, 1995) may be written, 
 denoting  perturbations 
by a prime: 
\begin{eqnarray}
{\partial v_r'\over \partial t}+i\Omega  v_r' -2\Omega v_{\phi}'& 
= & -{\partial W\over \partial r} 
 \nonumber \\
{\partial v_\phi'\over \partial t}+i\Omega v_\phi'
 + v_r'r^{-1} \frac{d(r^2 \Omega)}{dr} & = &  -\frac{i W}{r}
\nonumber \\
 {\partial v_z'\over \partial t}+i\Omega v_z'& = &
-\frac{\partial W}{\partial z}
\label{velocity}
\end{eqnarray}                  
 Here $W= P'/\rho = \rho' c_s^2 /  \rho,$ with $c_s$ being the 
sound speed.
The linearized continuity equation may be written as
\begin{equation}
{\rho \over c_s^2}\left({\partial W \over \partial t}+i\Omega W \right) =
-{1\over r}\frac{\partial (r \rho v_r')}{\partial r} 
-\frac{ i \rho v_\phi'}{r}
-\frac{\partial (\rho v_z')}{\partial z}. \label{lincont} \end{equation}

For a  disc in a  spherically symmetric 
external potential,  because there is no preferred direction for the
disc rotation axis,
equations (\ref{velocity}) and (\ref{lincont})
have  a  solution corresponding to 
a time independent rigid tilt. Then  $W= -irz \Omega^2 g,$
where $g$ is a constant inclination. The
corresponding components of the velocity perturbation are
 $v_z'= r\Omega g, v_r'= -z \Omega g,  v_\phi'= -izg(d(r\Omega) /dr).$
Here we recall that for a barotropic disc $\Omega$ is
a function of $r$ only.

For perturbations corresponding to large scale warps,
we expect the local inclination
 to vary  on  a  radial length scale significantly greater than the thickness 
$H.$ Then  it is reasonable to assume
that  it is also approximately independent of $z.$
We remark that this was found to be the case in  the  linear calculations
of bending waves by Papaloizou \& Lin (1995), which 
took the vertical structure of the disc fully into account,
even when the radial  wavelength was comparable to the vertical thickness.
Thus to describe bending waves with long radial
wavelength, we set
$W= -irz \Omega^2 g,$ and assume the local inclination
 $g$ and thus $v_z'$
are approximately independent
of $z.$ Then (\ref{velocity}) implies that $v_r',$ and $v_\phi'$
are both proportional to $z.$ Therefore  we set
$v_r' = z q_r',$ and $v_\phi'=z q_\phi' $ respectively, where
$q_r',$ and $q_\phi'$ are also assumed approximately
independent of $z.$ Equation (\ref{velocity}) then gives

\begin{eqnarray}
{\partial q_r'\over \partial t}+i\Omega  q_r' -2\Omega q_{\phi}'& 
= & i{\partial  r \Omega^2 g \over \partial r}
 \nonumber \\
{\partial q_\phi'\over \partial t}+i\Omega  q_\phi'
 + q_r'r^{-1} \frac{d(r^2 \Omega)}{dr} & = &  -\Omega^2 g
\nonumber \\
 {\partial v_z'\over \partial t}+i\Omega v_z'& = &
ir\Omega^2 g
\label{velocityz}
\end{eqnarray}           

The continuity equation (\ref{lincont}) is then
multiplied by $z$ and vertically integrated through the disc with
the result

\begin{equation}
{\cal I}r\Omega^2
\left({\partial g \over \partial t}+i\Omega g \right) =
-{i\over r}\frac{\partial (r \mu q_r')}{\partial r}
+\frac{ \mu q_\phi'}{r}
+i \Sigma v_z', \label{lincontz} \end{equation}  
where
\be {\cal I} = \int^{\infty}_{-\infty} \frac{\rho z^2}{c_s^2} dz, 
\label{I} \ee and 
\be \mu = \int^{\infty}_{-\infty} \rho z^2 dz. \label{mu} \ee

Equations (\ref{velocityz}) and (\ref{lincontz}) provide
a set of linear equations for the inclination, $g$, as a
function of $r$ and  $t.$ They may be used
to follow the evolution of a disturbance from initial data.
We compare the time dependent evolution calculated
in this way with that obtained from non-linear SPH calculations below.

\subsection{Low Frequency Bending Waves}

For the purpose of discussing  low frequency waves,
it is  convenient to consider  solutions  of the linearized equations
for which 
the  $t$ dependence is  separated through a factor   
$e^{i \sigma t},$ where $\sigma$ is the mode frequency.

The components of the  velocity perturbations 
are then given from (\ref{velocity})  by  (e.g. Papaloizou \&  Terquem 1995):

\begin{eqnarray}
\frac{v_r'}{z} & = & - (\sigma + \Omega) g + 
\frac{\left( r \Omega^2 (\sigma + \Omega) \frac{\partial g}{\partial r} +
g \sigma^2 (3 \Omega + \sigma) \right)}{(\sigma + \Omega)^2 - \kappa^2} \nonumber \\
\frac{v_{\phi}'}{z} &  = & \frac{i g \Omega^2 + i (v_r'/z)r^{-1} \frac{d(r^2 \Omega)}{dr}}{ (\sigma + \Omega)} \nonumber \\
v_z' & = & \frac{\Omega^2 r \frac{\partial (z g)}{\partial z}}{(\sigma + \Omega)} \label{velocitytp}
\end{eqnarray}
where 
\be \kappa^2 = \frac{2 \Omega}{r} \frac{d(r^2 \Omega)}{dr} \nonumber \ee
is the square of the epicyclic frequency.

For low frequency modes in a near Keplerian disc with $|\sigma| \ll \Omega,$
it should be noted that there
is a near--resonance
through the near vanishing of $(\sigma + \Omega)^2 - \kappa^2,$
with the result that
slowly varying warps may induce large horizontal motions and vertical shear
$\left(\frac{\partial v_r'}{\partial z}, \;
\frac{\partial v_{\phi}'}{\partial z} \right)$. When considering the effects of
viscous dissipation, provided the viscosity is not highly
anisotropic,  it is the damping of this vertical shear that provides
the dominant effect of viscosity.

The inclination, $g$, is governed by a 
single, second order ordinary  differential equation 
(e.g. Papaloizou \& Lin 1994, Papaloizou \& Terquem 1995).
For the low frequency limit we are interested in,  this may 
be approximated as:
\be 4 g \Omega (\omega_z + \sigma) {\cal I} + \frac{d}{dr} \left( {{ \mu 
\Omega}\over {\sigma + \Omega - \kappa}} \frac{dg}{dr} \right) =0. \label{g1}
\ee
Here  the free particle
nodal  precession frequency, $\omega_z,$  is given by
\be 2\Omega \omega_z =\left(\Omega^2 -  \frac{\partial^2 \Phi}{\partial z^2}\right).\ee

In a near Keplerian disc,  $\omega_z$ and the apsidal precession frequency
$\Omega -\kappa$ are small compared to $\Omega$  and  low frequency disturbances
may be considered for which 
$\sigma$ is of comparable  magnitude.

For a strictly Keplerian disc where $\omega_z=0$ and $\kappa=\Omega$, 
from (\ref{g1}) it may be seen that
 localised disturbances, for which $g \propto e^{i k r}$
obey the dispersion relation given by
\be \sigma^2 = a^2 k^2 \label{dispreln}\ee
where 
\be a= \frac{1}{2} \sqrt{\frac{\mu}{{\cal I}}} = \frac{\bar c_s}{2}. \ee
It may be seen from the definition of $\mu$ and $\cal I$ in equations (\ref{I})
and (\ref{mu}) that $\bar c_s$ is a  mean sound speed, such  that
in the limit $\sigma \rightarrow 0,$
warping disturbances propagate without  dispersion at half of this mean 
sound speed. For finite $\sigma,$ dispersive  corrections are of order $\sigma/\Omega.$

The full set of 
equations (\ref{velocityz}) and (\ref{lincontz}),
which allow for perturbations of arbitrary frequency,
are used to calculate the evolution of linear bending waves below.
However, the type of disturbances  considered can in fact be well described
using the low frequency approximation (see  Nelson, Papaloizou \& Terquem 1999)
and so may be regarded as effectively decoupled from other high frequency modes.

\subsection{Effect of a Small Viscosity} \label{Smallvisc}
In a Keplerian disc, the most important effect of a small viscosity
is to act on the resonantly induced horizontal  motions  through the 
($r$, $z$) and ($\phi$, $z$) components of the viscous tensor.
If the kinematic viscosity is taken to be 
\be \nu = \frac{\Gamma {\cal Q}}{\rho}, \label{nu} \ee
where $\Gamma$ is an arbitrary function of radius and ${\cal Q}=\int_z^{\infty}
\rho z dz$, then the main effect of the viscosity is to replace the 
resonant denominator 
$$ \sigma + \Omega - \kappa$$
in equation
(\ref{g1}) by 
$$ \sigma -i \Gamma + \Omega -\kappa.$$
In a thin Keplerian disc,
$$\int_z^{\infty}
\rho z dz = P/\Omega^2.$$
Setting $\Gamma= \alpha_1 \Omega,$ we find $ \nu =(\alpha_1 P)/(\rho \Omega)$
which gives a  prescription  
equivalent to
that of 
Shakura \& Sunyaev (1973). 
This scheme for incorporating viscosity is
included in  the governing equations (\ref{velocityz}) and (\ref{lincontz})
for the time dependent evolution of warping disturbances by
the straightforward  operator replacement
$$\left({\partial  \over \partial t}+i\Omega  \right)
\rightarrow  \left({\partial  \over \partial t}+i\Omega +\Gamma \right)$$
in equations (\ref{velocityz})  while leaving 
 (\ref{lincontz}) unaltered. 

However, it should be
noted that the  $\alpha_1$  here differs from the `$\alpha$'
parameter usually considered
in thin disc theory since it acts through the ($r$, $z$) and ($\phi$, $z$)
components of the viscous tensor rather than through the ($r$, $\phi$) 
component. For this reason we  introduce the subscript $1$ from now on.
 
When viscosity is introduced in this way, equation (\ref {g1}) is 
modified to read
\be 4 g \Omega (\omega_z + \sigma){\cal I} + \frac{d}{dr} \left( 
\frac{ \mu \Omega
 }{\sigma - i \alpha_1\Omega + \Omega - \kappa} \frac{dg}{dr}
\right) = 0 \label{gvisc1} \ee
The associated local dispersion relation (for $\omega_z = \Omega-\kappa =0$)
is  given 
through
\be \sigma = \left(\frac{a^2 k^2}{\sigma - i \alpha_1 \Omega} \right).
\label{vdispreln}\ee
If $|\sigma| << \alpha_1 \Omega$ in the low frequency limit,
then equation (\ref{vdispreln}) becomes
$$ \sigma = \frac{i a^2 k^2}{\alpha_1 \Omega}.$$
This is the form of the dispersion relation one
obtains from a diffusion equation
which implies that for a Keplerian disc, warps
evolve diffusively with diffusion coefficient  given by
\be {\cal D} = \frac{{\bar c_s^2}}{4 \alpha_1 \Omega}.  \ee
The corresponding time scale for global evolution of
the disc is given by
\be  t_{diff}= {r^2 \over {\cal D}} . \label{t-diff} \ee

The transition between  wave--like and diffusive regimes is expected 
to occur when the diffusion time above,  calculated  for the whole disc,
becomes comparable to the wave propagation time across it, or when 
 $\alpha_1  \sim H/r$ (e.g. Papaloizou \& Lin 1994).

It is very important to note that provided $\alpha_1$ is
significantly below unity,  in either regime
the communication of warp  information across the disc
can be very much more rapid than  
the time scale for matter to accrete through it.
 This is in fact confirmed in our simulations and expected
from Papaloizou \& Pringle (1983).

The non linear  numerical simulations  of polytropic discs
presented in this paper apply for the most part 
to  the parameter regime in which $\alpha_1 \; \ls \; H/r$, so that 
the time dependent behaviour
of small warping disturbances should exhibit wave--like behaviour. To
check that this is indeed the case, we have performed a number of
calculations of the evolution of small
amplitude bending waves in section (\ref{waves}).  
Results obtained with SPH  are  compared
directly with the predictions of linear theory given by integrating
equations (\ref{velocityz}) and (\ref {lincontz})
forward in time,
starting from the appropriate initial conditions, and modified to include the
effects of viscosity in the manner described in section (\ref{Smallvisc}).

\begin{figure*}
\epsfig{file=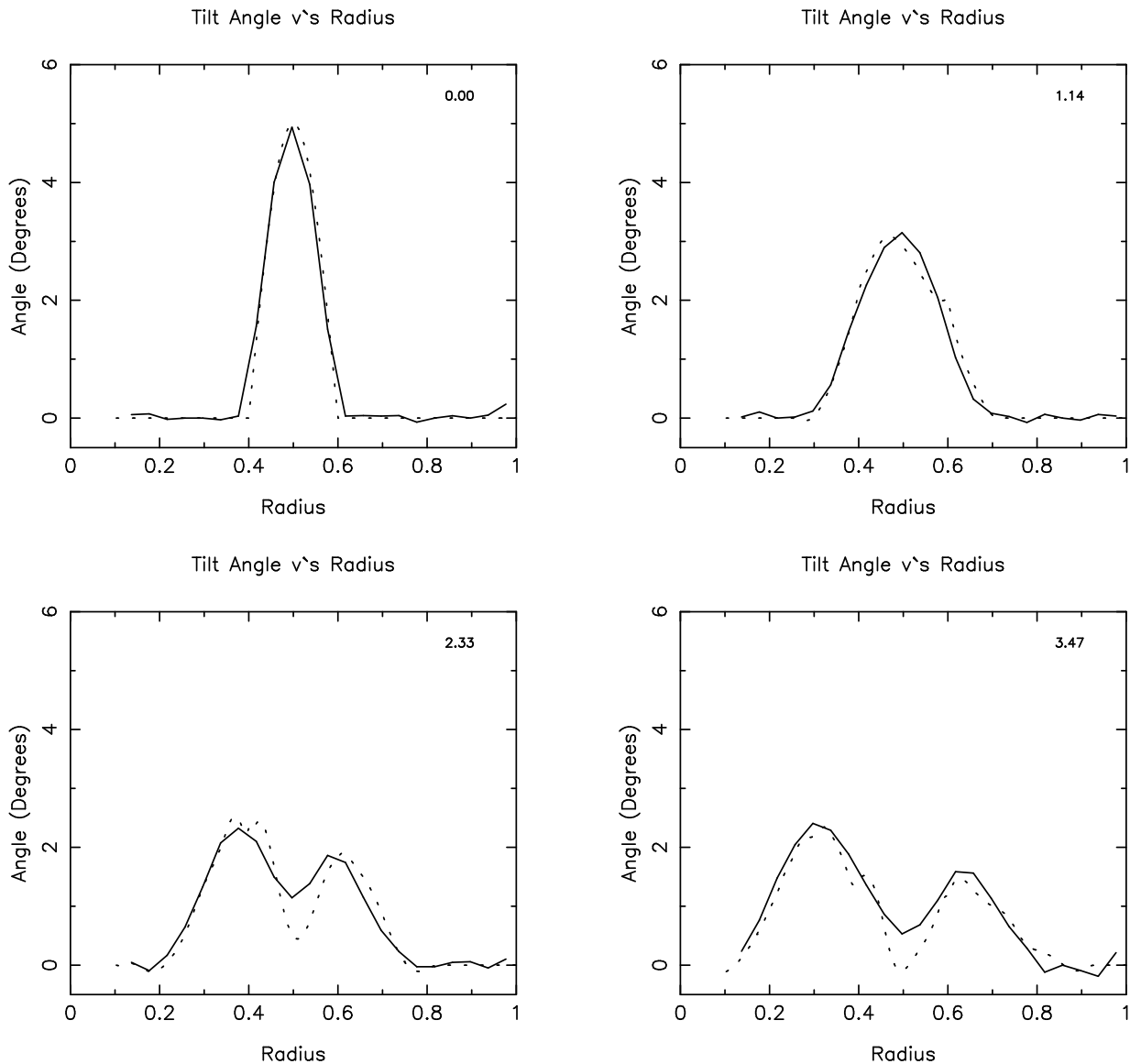, width=\textwidth}
\caption{This figure shows the variation
of disc tilt versus radius, as a
function of time, for a linear calculation ({\em dotted} line),
and a non linear SPH calculation ({\em solid} line).
The assumed parameters for the linear calculation were
${\cal M} =8.33$, $\alpha_1=0.04/r^{1/2}$. The SPH calculation is described
in the text.}
\end{figure*}

\begin{figure*}
\epsfig{file=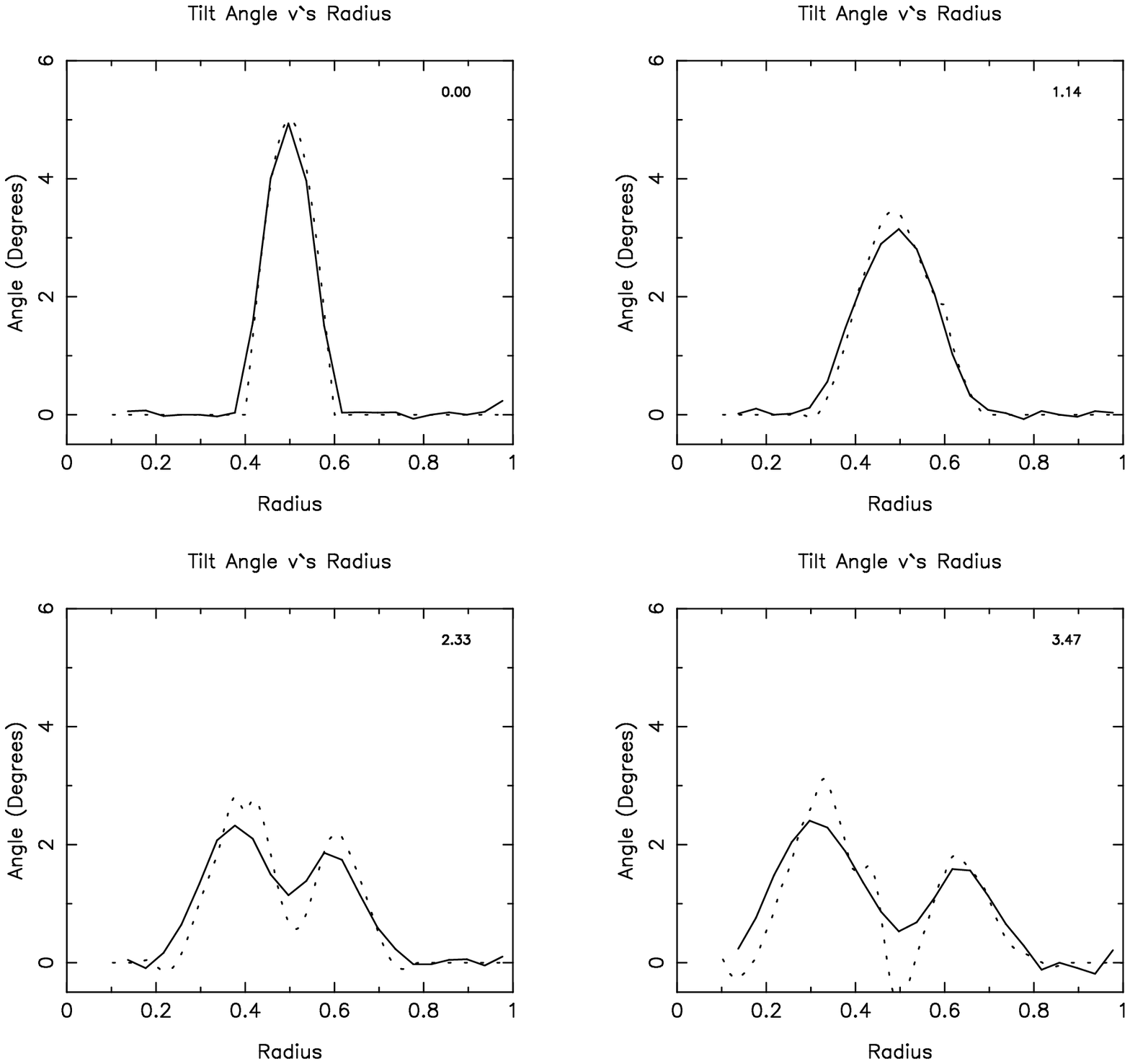, width=\textwidth}
\caption{This figure shows the variation of
disc tilt versus radius, as a
function of time, for a linear calculation ({\em dotted} line),
and a non linear SPH calculation ({\em solid} line).
The assumed parameters for the linear calculation were
${\cal M} =10$, $\alpha_1=0.0$. The SPH calculation is described
in the text.}
\end{figure*}

\section{Numerical Method} \label{Num}
The set of fluid equations described in section (\ref{basic-eq})
are solved using smoothed particle hydrodynamics
(Lucy 1977; Gingold \& Monaghan 1977). SPH uses particles to represent a subset
of
the fluid elements that arise in the Lagrangian description of a fluid.
The version of SPH used in the calculations presented here is a conservative
formulation of the method that employs variable smoothing lengths
(Nelson \& Papaloizou 1994). For the sake of brevity, we do not provide a
detailed description of the method, but only highlight those points salient
to the work presented in this paper. A detailed description of the code
is presented in Nelson \& Papaloizou (1994), along with a number of
test calculations.

In order for the method to be conservative, the smoothing lengths must be
functions only of the interparticle separations. Accordingly, we find the
${\cal N}_{TOL}$ nearest neighbours and calculate the smoothing lengths, $h_i$,
using the expression
\begin{equation}
h_i = \frac{1}{N_{far}} \sum_{n=1}^{N_{far}} \frac{1}{2}
| {\bf r}_i - {\bf r}_n |
\label{hn6}
\end{equation}
where the summation is over the $N_{far}$ most distant nearest neighbours of
particle $i$. For the calculations presented here, we take
${\cal N}_{TOL}=45$ and $N_{far}=6$.

The numerical method employs an artificial viscosity term which allows shocks
to be properly modeled. The magnitude of this viscosity is controlled by
two parameters, $\alpha_{SPH}$ and $\beta_{SPH}$ as defined by 
Nelson \& Papaloizou (1994).
We set these parameters to have the values $\alpha_{SPH}=0.5$ and
 $\beta_{SPH}=0.0.$

It should be remarked  that in addition to providing a bulk viscosity,
the artificial viscosity provides a shear viscosity arising from the finite
resolution of the method which allows particles to communicate across
shearing interfaces. The particles also develop random velocities
that result from stochastic fluctuations in their pressure forces, and
these motions lead to a diffusive evolution of the particle distribution.
Test calculations that used ring spreading simulations 
(Larwood {\em et al} 1996), and more recent calibration runs 
(Bryden {\em et al.} 1999) indicate that these processes lead to the viscous
evolution of accretion disc models calculated with the SPH code. The
code viscosity is characterised by
a range of
values for the dimensionless shear viscosity parameter,
$\alpha$, that depend on the value of $H/r$ in the disc. The dimensionless
parameter $\alpha$ is defined through the expression for the kinematic
viscosity $\nu = \alpha H^2 \Omega$,
where $H/r= {\cal M}^{-1}$,  with  ${\cal M}$  being
the midplane Mach number in the
disc. This does not vary greatly over the radial extent
of  most of the models considered here.
 For disc models with
${\cal M} =10$ ({\em i.e.} with aspect ratio $H/r \sim 0.1$) we find that
$\alpha \simeq 0.02$, and for models with ${\cal M} =30$ we find that
$\alpha \sim 0.03$ -- $0.04$.

\section{Bending Wave Calculations} \label{waves}
We have performed a number of calculations in which small amplitude, localised 
warping disturbances were applied to accretion discs, and then allowed
to evolve. The results of these calculations were compared
with the results  obtained from the linear theory 
 through equations (\ref{velocityz}) and (\ref{lincontz})
discussed in section (\ref{linear}).
These
simulations, and their comparison with linear calculations
which used a surface density profile matching that of the discs, 
are presented below.

\begin{figure*}
\epsfig{file=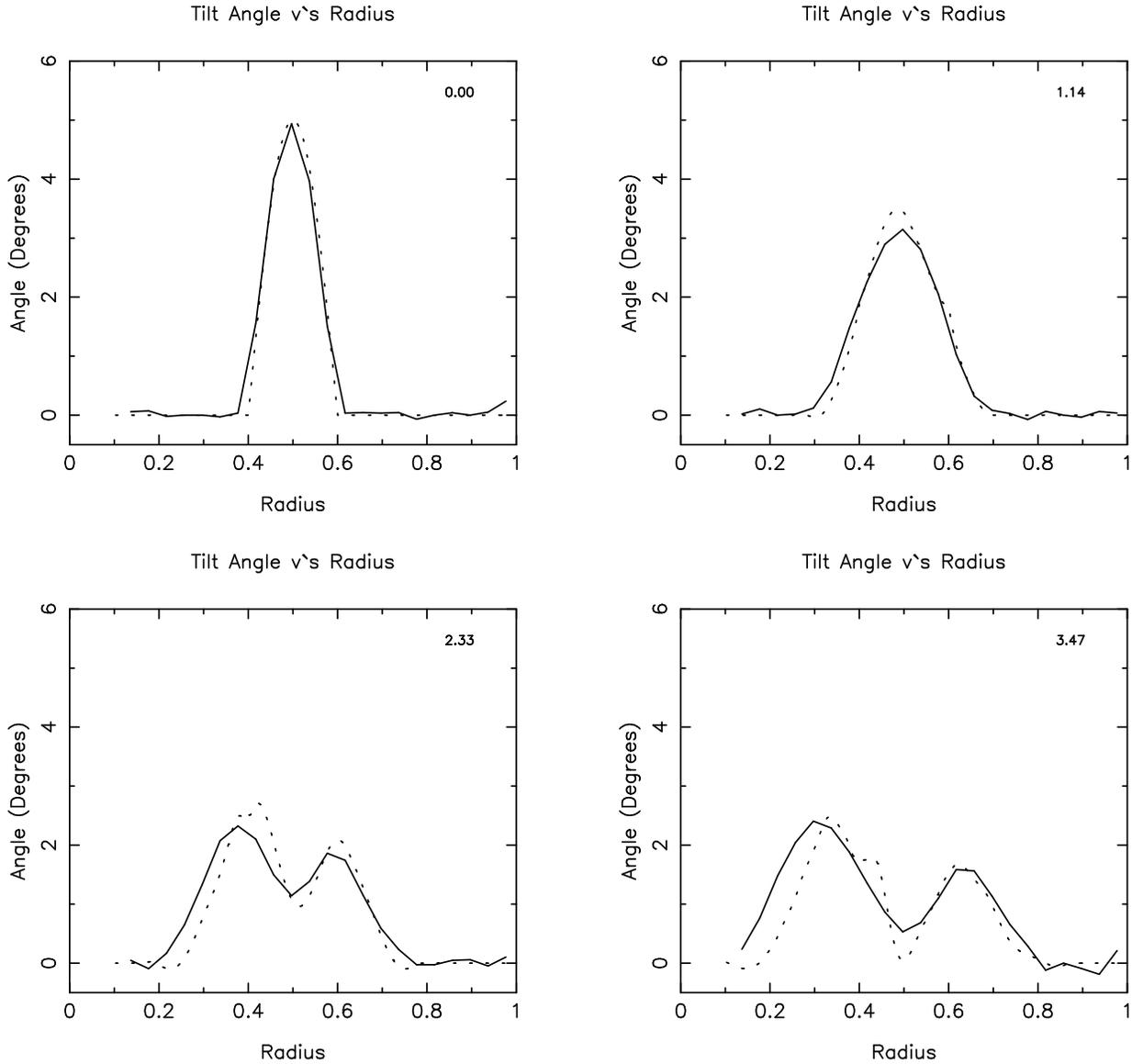, width=\textwidth}
\caption{This figure shows the variation of
disc tilt versus radius, as a
function of time, for a linear calculation ({\em dotted} line),
and a non linear SPH calculation ({\em solid} line).
The assumed parameters for the linear calculation were
${\cal M} =10$, $\alpha_1=0.04/r^{1/2}$. The SPH calculation is described
in the text.}
\end{figure*}

\begin{figure*}
\epsfig{file=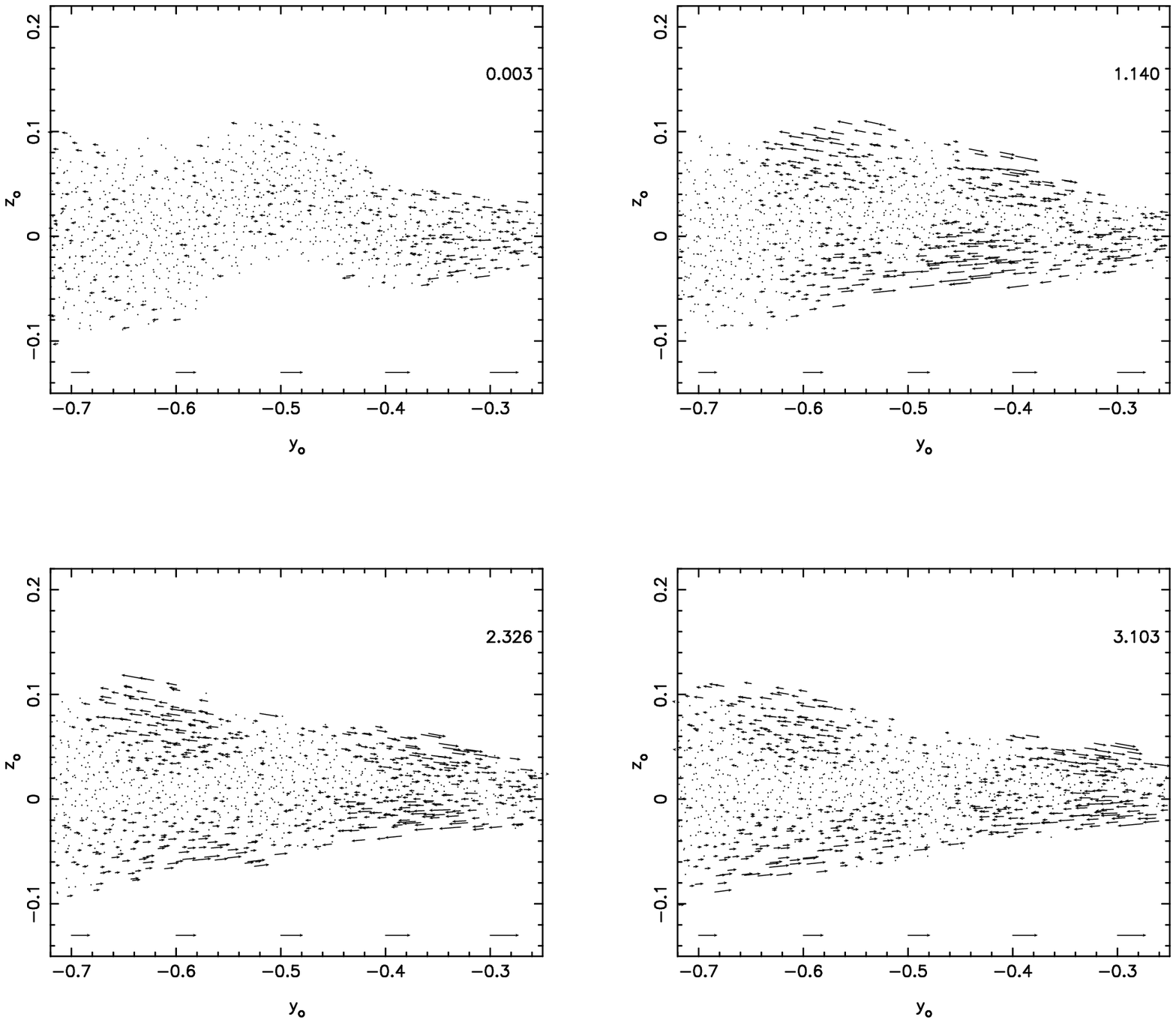, width=\textwidth}
\caption{This figure shows the perturbed radial velocity arising from
bending wave propagation through the disc. Note the vertical shear
as perturbed radial motion, which is odd in $z$, is generated in the disc
by the warp. The arrows at the bottom of each panel indicate the
magnitude of the midplane sound speed at each radius.}
\end{figure*}

\begin{figure*}
\epsfig{file=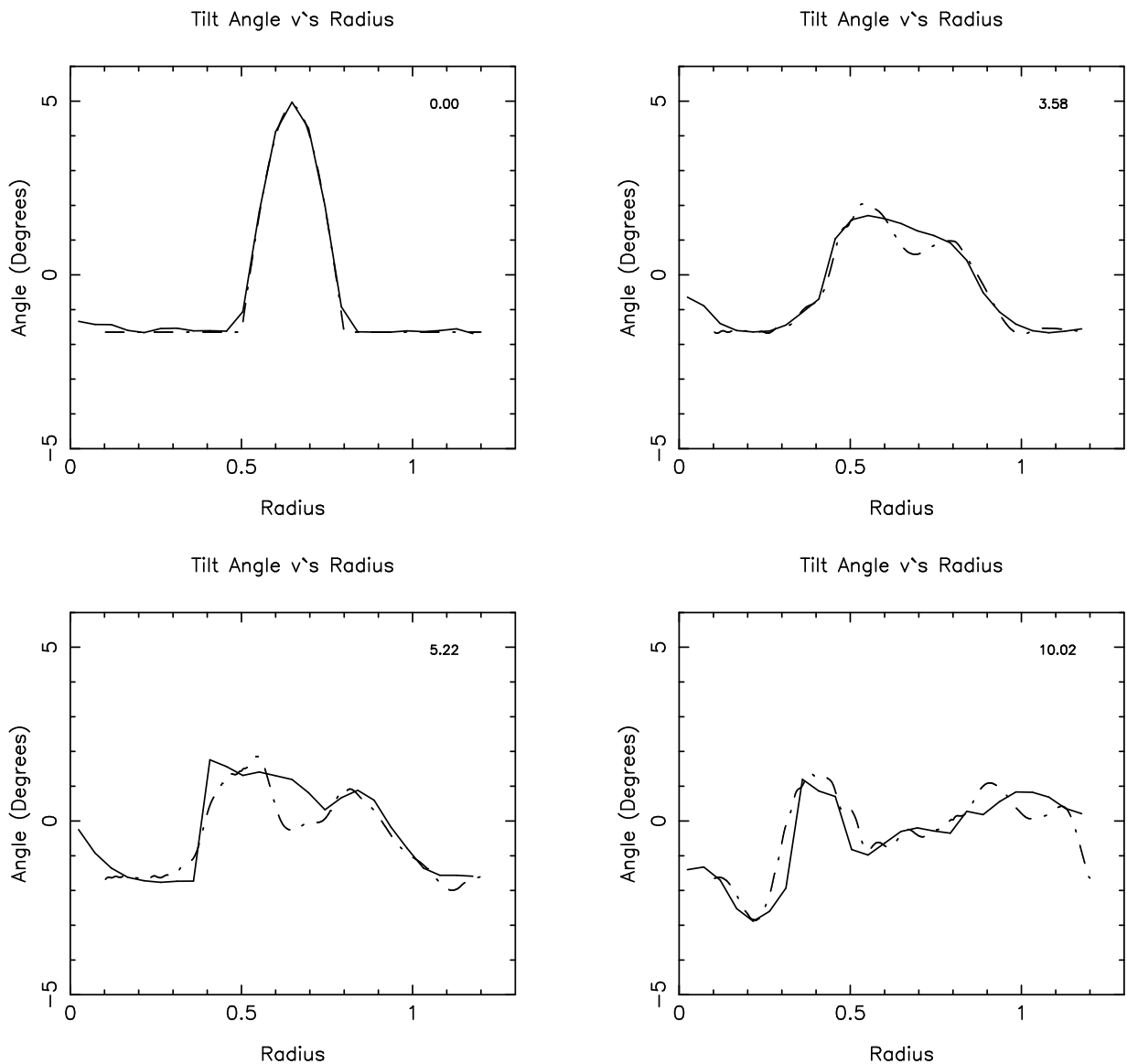, width=\textwidth}
\caption{This figure shows the radial variation of tilt angle,
as a function
of time, for a bending wave calculation performed for a disc orbiting in
a softened potential. The {\em solid} line shows the results from the
non linear SPH calculation, which is described in the text.
The {\em dot--dashed} line shows the evolution of a
linear calculation, with assumed parameters
${\cal M}=10$, $\alpha_1=0.04/r^{1/2}$.}
\end{figure*}

\begin{figure*}
\epsfig{file=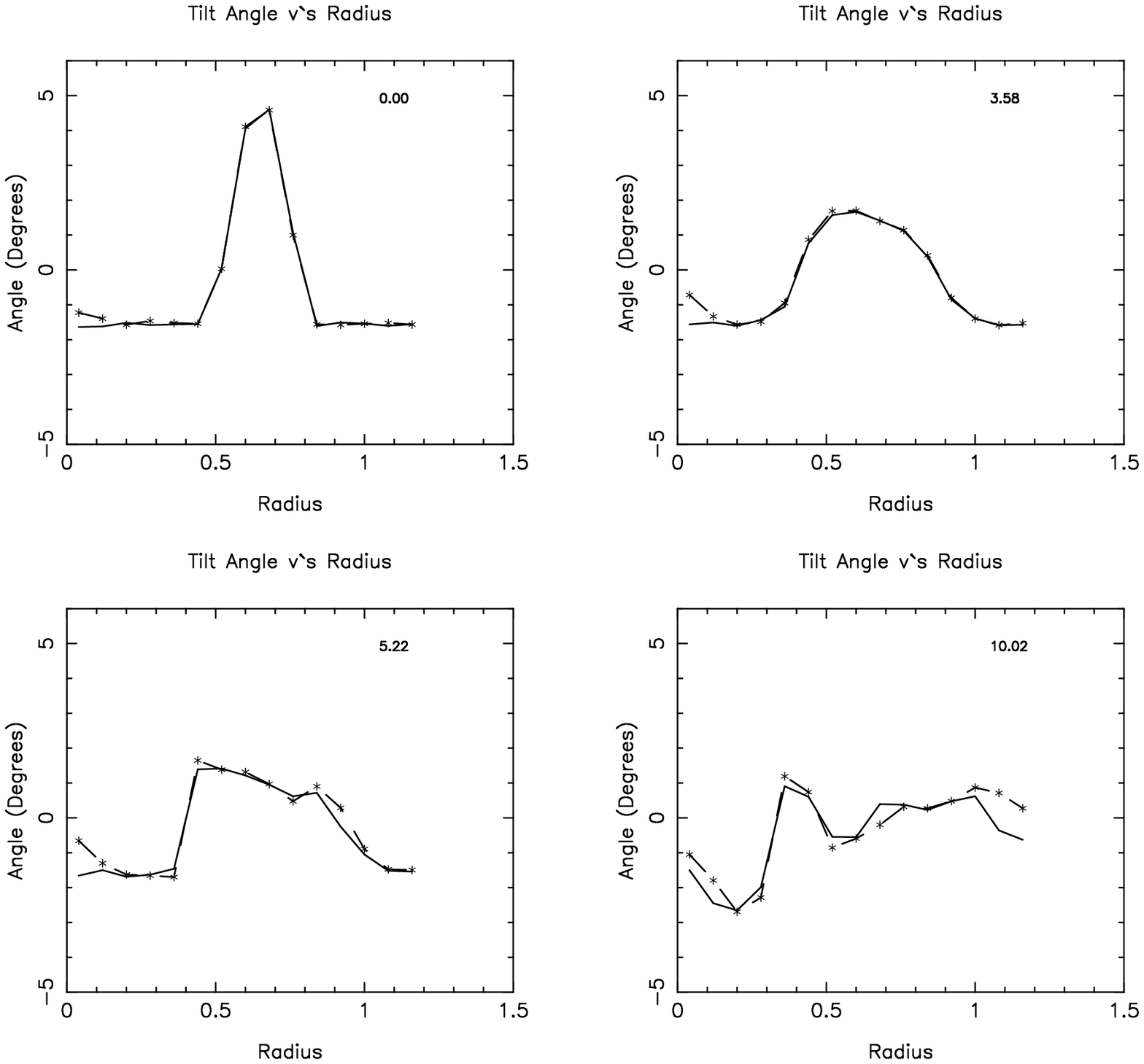, width=\textwidth}
\caption{This figure shows the radial variation of tilt angle,
as a function
of time, for two bending wave calculations performed with discs orbiting in
softened potentials. The {\em solid} line shows the results from a
non linear SPH calculation using 20,000 particles, the {\em dashed} line
shows results from a calculation employing 102,000 particles.}
\end{figure*}

\subsection{Initial and Boundary Conditions}
The calculations presented in this section employed differing
numbers of particles, and were performed for two different values of
the gravitational softening parameter, $b$, introduced in section
(\ref{basic-eq}). 
The central mass in each of these calculations was taken to be $M=1$,
with the gravitational constant $G=1$.
The initial conditions for each calculation performed are described below.

A number of calculations were performed for discs orbiting in 
purely Keplerian potentials, in which the softening parameter
$b=0$. In these cases, 
the disc models were constructed by creating a gaseous annulus, 
with a free inner boundary at $r=r_{in}=0.2$ and a free 
outer boundary at $r=r_{out}=0.8$,
rotating about the $z$ axis.
This annulus contained $N=102,000$ particles, and the required disc thickness
(or equivalently disc midplane Mach number)
was obtained by setting the polytropic constant, $K$, to the appropriate value.
These disc models were then allowed to relax by evolving them for a 
time that was slightly greater than two orbital periods at their outside edge.
The surface density was initially constant, and was found to evolve very little
during this process of relaxation.

The bending wave calculations for these disc models were initiated
by tilting an annulus in the disc, which 
lay between $r=r_{min}=0.4$ and $r=r_{max}=0.6$. The tilt was
imposed by rotating about  the diameter coincident
with the  $x$ axis. 
The tilt amplitude varied in radius  according to the expression
\be g(r)=g_{0} \sin\left[\frac{\pi (r-r_{min})}{\Delta r}\right] \label{g} \ee
where
$\Delta r = r_{max}-r_{min}$.
Calculations were performed for discs in which the midplane Mach number
was ${\cal M} =10$ and $g_0$ took the values of 5, 10, and 20 degrees.
An additional calculation was performed in which the midplane Mach number 
${\cal M} =30$ and $g_0=5$ degrees.

A suite of calculations were performed in which the discs 
extended all the way to $R=0$, so that softening of the central potential was
required. In these cases, units were chosen
such that  the gravitational softening parameter $b=0.2$,
and the disc  outer  radius $R=1$. Calculations in this case
used differing number of particles, with $N=20000$, $40000$, and $102,000$.
The disc thickness (or mid plane Mach number) was obtained by setting
an appropriate value of $K$, and the initial disc tilt was 
obtained using equation ({\ref{g}), with $r_{min}=0.5$ and $r_{max}=0.8$.
Tilt amplitudes with $g_0=$ 3, 7, 10, and 15 degrees were considered, with
the midplane Mach number being ${\cal M}=10$.
When discussing the results of these calculations below, we will
concentrate primarily on the cases with $g_0=7$ degrees, and 
$N=20000$ and $102,000$ particles, in order to compare the effects of
changing the particle number and resolution on the results.

\subsection{Results}
We now present and compare the results from a subset of the free bending
wave calculations. We first discuss the results for a low 
amplitude ($g_0=5^o$) warping disturbance applied to a Keplerian disc.
We also present results for two calculations performed with discs
orbiting in softened potentials, but in which the particle numbers used
differed from one another. Finally, we compare the results of three
calculations in which the initial tilt amplitudes differed from one another,
focusing primarily on a low amplitude tilt and a high amplitude tilt
calculation.

\subsubsection{Low Amplitude Tilt in Keplerian Disc} \label{linkep}
The plots presented in Fig. (1) show the time evolution of the disc tilt
$g(r)$ during a calculation
performed for a disc with ${\cal M}=10$ and $N=102000$ particles. 
When discussing the figures in this paper, we will use the convention
that the top left panel will be referred to as panel 1, with
the remaining panels being labelled as 2, 3, and 4 when moving from left to
right and from top to bottom. The time corresponding to each panel is shown
in the top right hand corner in units of $\Omega^{-1}$ evaluated at $R=1$.
The initial
tilt applied to the disc was of the form given by equation (\ref{g}), with
$g_0=5^o$, and may be seen in panel 1 of Fig. (1). The solid line in Fig. (1)
represents the variation of tilt angle $g$ with $R$ for the non linear SPH
calculation, whereas the dashed line represents a  linear calculation
computed using the  time dependent linearized equations 
described in section (\ref{linear}).

Moving from left to right and from top to bottom, we can observe that the
initial `pulse' starts to broaden as the locally applied disc warp begins
to propagate in towards the disc centre and out towards its exterior edge.
After a time of $t=2.33$, the initial pulse can be seen to be splitting into two
separate pulses, corresponding to an in--going and an out--going bending wave,
 as
one would expect from linear theory. After a time of 
$t=3.47$ the waves have more or 
less reached the outer edge of the disc, and are almost fully separated.
The parameters used in fitting the linear calculation to the
non linear SPH calculation assumed a disc model in which the midplane
Mach number ${\cal M}=8.33$, and took a variable value of  the 
viscosity parameter
$\alpha_1 = 0.04/r^{1/2} $,
where this  radial dependence of the viscosity
arises from assuming an isotropic viscosity, and a steady state
disc with uniform
surface density $\Sigma$, aspect ratio $(H/r)$, and mass flux.

In addition to the fit presented in Fig. (1), we also provide a comparison
between the nonlinear SPH calculation and two further linear
calculations in Figs. (2) and (3). The linear calculation in Fig. (2)
assumed a disc midplane Mach number ${\cal M} = 10$, and neglected
the effects of viscosity entirely ({\em i.e} $\alpha_1 =0.0$).
It is apparent that the SPH calculation mimics the broad features 
of the linear calculation in a general sense, but that a good fit is
not obtained since viscosity is required to damp the amplitude
of the waves. The linear calculation presented in Fig. (3) assumed
${\cal M}=10$ and $\alpha_1=0.04/r^{1/2}$.
In this case
it appears that the inward propagating pulse produced by the SPH
calculation travels too fast, and  the leading edge
is not particularly well fitted by
the linear calculation presented. The outward moving pulses in both the linear
and nonlinear runs appear to match very closely, however, as
shown in panel 4 of Fig. (3). We remark that we expect the results of the
SPH calculations to be more accurate for the outward moving bending
waves since the resolution is better in the outer parts of the disc models
than in the inner parts. 

All of the linear calculations presented in
Figs. (1) -- (3) provide reasonable fits to the nonlinear SPH calculation.
The region in which the largest discrepancy arises 
in the Figs. (1) and (2) cases is at a radius of
$R \sim 0.5$ -- {\em i.e.} the radial position about which the initial warping
disturbance was centred. The linear calculations predict that the tilt
angle, $g$, should drop sharply in this region as the 
in--going and out--going waves
separate, with the magnitude of this decrease in $g$ being determined
by $\alpha_1$. A larger viscosity leads to a more diffusive evolution
of the warp, and thus to pulses that separate less cleanly.
 It is possible that the effective viscosity ({\em i.e.} $\alpha_1$)
in the SPH calculations is a function of the local conditions,
and that the particle relaxation in the region initially
occupied by the applied warping disturbance
leads to a larger effective viscosity there which is
not accounted for in the simple  steady state
form of the viscosity  adopted in the
linear calculations. In our experience a very precise fit to the
simulations may be obtained if  an additional radial
dependence is allowed in $\alpha_1$ and ${\cal{M}}.$
However, we do not pursue this here.

The form of the perturbed velocity field expected to be 
induced in the disc when it is warped is given by equation (\ref{velocity}).
It may be seen that as a bending wave travels through the disc, it leads to the
excitation of radial motions that are odd functions of $z$, such that
a vertical shear is induced. This shearing motion may be observed
in Fig. (4). Each of the panels shows particle projections onto the
$y$--$z$ plane for a thin cross section of the disc centred on the $y$ axis.
The disc is  shown  for $-0.8 \le y \le -0.2 .$
The radial velocities ($v_r$) of the particles are indicated by an arrow,
but only for particles for which $|v_r|$ is above a  small threshold value.
The length of the arrow indicates the magnitude of the velocity, and the
arrows located at the bottom of each panel show the magnitude of the
midplane sound speed at
each radial position in the disc.
The top panel shows the initial state of the disc, with the applied warping
disturbance centred on $y=-0.5$ being apparent. The residual velocities
in this panel are just due to the residual random motions of the particles.
The middle panel shows the disc after it has evolved for $t=1.14$, and it is
obvious that  vertical shearing motions have been excited by the 
initial warp.
After  further evolution, the initial pulse separates into an inward and outward 
travelling bending wave, leaving the structure of the disc at the position of
the initial pulse relatively unchanged. The progression of these two pulses may
be seen in the final two panels of Fig. (4), where two distinct waves are visible, 
separated by a `dead--zone' at $y \sim -0.5$. This figure illustrates the fact
that the code's performance is such that it is able to capture the essential
features of bending wave propagation through Keplerian accretion discs, 
indicating
that the vertical structure of the disc models are approximated to adequate
accuracy.

In addition to performing bending wave calculations for ${\cal M}=10$
discs, a calculation for a ${\cal M}=30$ disc was also performed,
with $g_0=5^o$ in equation (\ref{g}). In this case, the value of the
dimensionless viscosity coefficient $\alpha$ is expected to be
larger than the disc aspect ratio, since calibration experiments indicate
that $\alpha \simeq 0.04$. These calibration
calculations measure the value of
$\alpha$ acting through the ($r$, $\phi$) component of the viscosity
tensor, rather than that which acts on the vertical shear, but should 
nonetheless
provide an estimate of the $\alpha_1$ acting on this vertical shear.
It is expected that a warping disturbance in a disc with
$\alpha_1 > H/r$ will evolve diffusively, rather than in a wave--like manner,
and this is what is observed. The non linear SPH calculation was fitted using
the equations from linear theory, and the best fit indicated
that $\alpha_1 \sim 0.2$ for an assumed value of the midplane
Mach number ${\cal M}=30$. There are probably two
reasons why this somewhat large value 
is found.

Firstly the initial pulse is non linear in this case
because $g_0$  is about three times larger than $H/R\sim {1\over {\cal{M}}}$
in this case, whereas it is comparable when  $ {\cal{M}}= 10.$ 
It thus corresponds to $g_0=15^o$ for the latter Mach number. For
such non linear warps the indication is that the evolution is more
diffusive (see below) so resulting in a larger measured value for $\alpha_1.$
In addition, the vertical structure in the inner parts
of the disc in this calculation is not as accurately modeled as in the
${\cal M}=10$ calculation, because the smoothing lengths are on the order
of the disc thickness. This has the effect of weakening the pressure
forces that arise when the disc midplanes at adjacent radii are
misaligned due to  warping, and thus reducing the efficacy of   
both diffusive and wave--like
communication. This increases the 
apparent magnitude of 
$\alpha_1$ above that expected from calibration experiments that
measure the ($r$, $\phi$) component of viscosity.

\begin{figure}
\epsfig{file=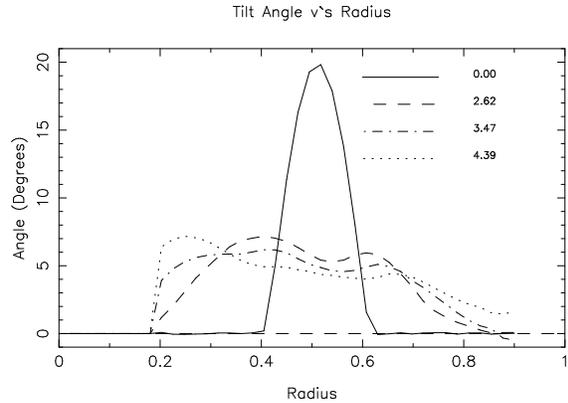, width=\columnwidth}
\caption{This figure shows the tilt angle versus radius,
as a function of
time, during a non linear SPH calculation in which the initial tilt
$g_0=20$ degrees. Note the non wave--like evolution compared with
previous, low amplitude runs.}
\end{figure}

\begin{figure*}
\epsfig{file=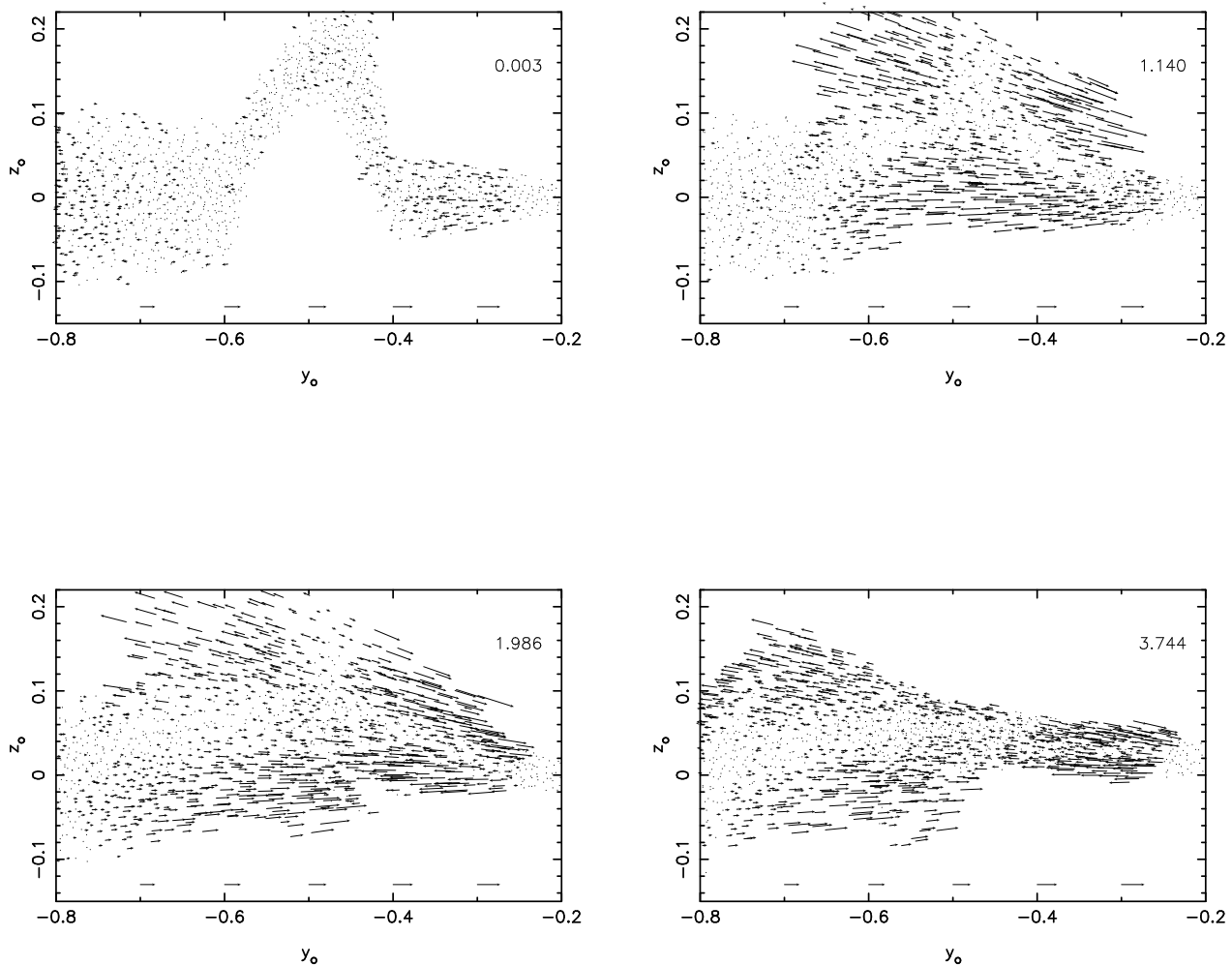, width=\textwidth}
\caption{This figure shows the perturbed radial velocity
arising from
bending wave propagation through the disc with $g_0=20$ degrees.
Note the vertical shear
as perturbed radial motion, which is odd in $z$, is generated in the disc
by the warp. The arrows at the bottom of each panel indicate the
magnitude of the midplane sound speed at each radius. It is apparent that sonic
radial motion is generated in the disc by the initial large amplitude warp.}
\end{figure*}

\subsubsection{Low Amplitude Tilt in Discs Orbiting in a Softened Potential}
A number of calculations were performed for discs which were orbiting
in softened Keplerian potentials. The discs in these calculations 
initially had an outer radius $R=1$, and extended in to $R=0$.
The gravitational softening length 
was $b=0.2$. Here we describe two calculations in which
${\cal M}=10$ and the tilt amplitude was $g_0=7^o$. One calculation 
employed $N=20,000$ particles, whereas the other employed $N=102,000$,
so that the purpose of this comparison is to examine how the
results vary as a function of particle number. In addition, we are 
able to observe the effects of a small departure from a Keplerian
rotation profile induced by the gravitational softening.

The results of the calculation employing $N=102,000$ particles
are plotted in Fig. (5) along with the results from a linear calculation.
It may be seen from these results the substantial effect of changing
from a Keplerian potential to a softened potential. 
The linear calculation 
(represented by the {\em dot-dashed} line) 
adopted a Mach number ${\cal M}= 10$ and  $\alpha_1=0.04/r^{1/2}$.
The {\em solid} line shows the 
evolution of $g(r)$ in the  SPH calculation. 
The fact that the  initial pulse  separates less  cleanly
into an inward and outward moving disturbance indicates 
a greater  and dominant dispersion resulting from the non zero value of $\Omega - \kappa$ 
in this softened potential case. The relatively poor fits at radii $R > 1.1$
at the latest time are probably due to the effects of a pressure
expansion at the outer edge during the SPH calculation,
which produced a low density region there. As in the previous example of a
disc orbiting in a Keplerian potential, the drop in the magnitude
$g$ around the position of the initial pulse is not well reproduced 
by the SPH calculation, and this may be again due to a localised 
larger effective
viscosity acting in the SPH run. Overall, however, the broad features
of the solution obtained using the linearised equations are
reproduced by the SPH calculation, particularly in the inner parts of the disc,
indicating that bending wave
propagation may be modeled using SPH with reasonable accuracy.

In Fig. (6), we plot the results from the two non linear SPH
calculations with different numbers of particles. Moving through the panels it is apparent that
the calculations are very similar in terms of the time evolution of the tilt,
and indicate that accurate results may be obtained with particle numbers
as low as $N=20,000$ for these particular calculations.  This is because
the vertical structure is well enough resolved over most of the
radial extent of the discs for ${\cal M}=10$. 

\subsubsection{Evolution of Nonlinear Warps}
In order to examine the effect of increasing the amplitude
of the initial warping disturbance on the characteristics
of wave propagation, a number of calculations were
performed in which the initial warp amplitude was varied.
The calculations were performed with discs orbiting in Keplerian
potentials, whose midplane Mach number was ${\cal M}=10$, and which
employed $N=102,000$ particles ({\em i.e.} 
identical to the discs described in section
[\ref{linkep}]). The values of $g_0$ in equation (\ref{g}) took the values
of $5^o$, $10^o$, and $20^o$. The calculation for which $g_0 = 5^o$
was described in section (\ref{linkep}). Although not described here
in detail, the calculation performed with $g_0= 10^o$ showed behaviour
intermediate between the $g_0=5^o$ and $g_0=20^o$ cases, as expected.
The evolution of the tilt
angle $g(r)$ as a function of time is displayed in Fig. (7) for the calculation
with $g_0=20^o$. Comparing this figure with Fig. (1) shows the difference
between the $g_0 =5^o$ case, in which the warp  behaves like
a linear perturbation, 
and the case with $g_0 =20^o$, in which the warp is a non linear perturbation.
In particular, it may be seen that the evolution of a 
free ({\em i.e.} unforced),
non linear warping
disturbance is much less wave--like than for the linear disturbance, indicating
that non linearity in the system leads to enhanced dissipation of the
bending waves. The evolution of the velocity field for the calculation in which
$g_0=20^o$ is shown in Fig. (8), and should be compared with Fig.(4)
which shows the same for a linear wave ({\em i.e.} $g_0=5^o$). It is obvious
that the strong misalignment of the disc midplane arising from the $g_0=20^o$ 
warp leads to  a flow  convergence   with radial velocities 
 on the order of the sound speed, such that a shock forms 
at $r \sim 0.5$. This shock does not appear in Fig. (4) for 
which $g_0=5^o$, because
the degree of warping is insufficient to generate large enough
 radial motions in
the disc.
The existence of this shock, which can be more clearly observed
in Fig. (9), leads to a strong dissipation of the bending wave, such that
its temporal evolution  appears more diffusive than is the case
with the $g_0=5^o$ pulse. 
We note that the existence of the shock in this calculation arises
because of the form of the initial pulse applied to the disc, which 
automatically leads to  convergence  of the radial flow in the disc.

In a situation where the warping of the disc is monotonic rather
than being in the form of a localised pulse, there would be no such
convergence point, and presumably  this type of shock would not occur.
 However, if
the degree of local warping of the disc is such that radial
velocities on the order of the sound speed are generated
({\em i.e.} when the local curvature $r |\frac{dg}{dr}| \ge H/r$), 
then bending waves are
unable to sustain themselves since the disc material is induced
to move faster in the radial direction than the
wave can propagate. It is likely that such a situation will also lead to
enhanced dissipation in the system, such that the non linearity
of the warp will be damped.

\begin{figure}
\centerline{\epsfig{file=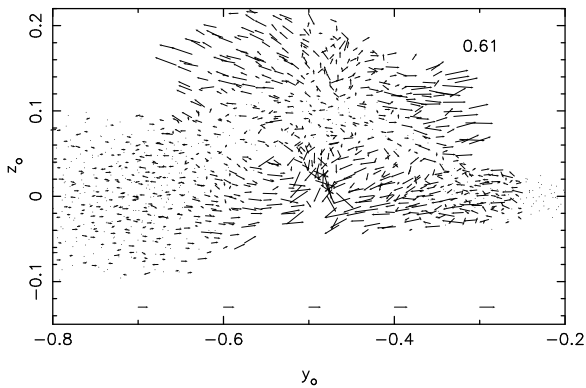, width=110mm}}
\caption{This figure shows the full velocity field for the high
 amplitude
 bending wave calculation. Note the convergence point at $y = -0.5$,
 corresponding to a shock.}
\end{figure}

\section{Discussion and Conclusion} \label{concl}
In this paper we have performed nonlinear simulations of bending waves
in accretion discs. Disc Mach numbers ranging between 10 and 30 have been
considered with particle numbers ranging between 20000 and 102000. The
main purpose of these calculations is to examine the ability of SPH to model
warped accretions discs, a problem which is intrinsically three dimensional in
nature.

Linear bending waves are expected to propagate if the $\alpha$ viscosity
appropriate to dissipation of vertical shear, $\alpha_1$ is $< H/r.$
In our simulations $\alpha_1\sim 0.04$ for discs with ${\cal M}=10$, so bending waves should propagate in this case. 
We performed
tests that showed that the linear wave propagation found in the simulations
could be well fitted by solutions of the linearized  problem
found  using a  finite
difference scheme.
This is true
for  strictly Keplerian and slightly non Keplerian discs, where
the major difference observed between these two cases is a more dispersive 
evolution of the bending waves when the central potential is non Keplerian.

A transition from wave--like to diffusive propagation of warps is expected to 
occur when $\alpha_1 > H/r$. Such a transition in behaviour was observed
in the non linear simulations when the disc Mach number was increased to 
${\cal M}=30$, since we estimate that $\alpha_1 > H/r$ in this case. 
We therefore remark that SPH simulations can model warped discs in both the
wave--like and diffusive regimes.

A detailed examination of the velocity field produced by the simulations
showed that the vertical shear motion predicted by Papaloizou \& Pringle (1983)
was reproduced, indicating that the vertical structure of the discs 
was  adequately modeled. 

Calculations of non linear warping disturbances were performed, where the
condition for non linearity is that the local curvature
$r |\frac{dg}{dr}| \ge H/r$. It was observed that non linear warps lead to an
increased dissipation due to shocks, restricting the perturbed horizontal 
motions to be subsonic. This was associated with a transition 
from wave--like to more
diffusive--like propagation of the warps, which has the same effect as 
would be produced by 
increasing the effective value of $\alpha_1$. 
We speculate that the effects of 
non linearity will always be to inhibit communication 
between different radial locations in the disc which
require  horizontal velocities that are supersonic
for the communication to be maintained.
We expect the effects of non linear
damping  to
adjust the warp amplitude until it becomes a linear perturbation.
This behaviour is similar to that produced
by an increase of the effective value of $\alpha_1$.

The simulations that we have presented in this paper indicate that SPH is able
to model warped accretion discs in the regime where warps are expected to 
propagate as bending waves, and also in the regime where the propagation is
diffusive.  The modeling of warped discs represents a stringent test of any
hydrodynamical code, since it requires the accurate modeling of
complex motion in three dimensions. We find that accurate results may be 
obtained even for rather modest numbers of particles
({\em i.e.} $N \sim 20000$). We have previously used SPH simulations to
examine the structure of warped accretion discs in misaligned binary systems 
(Larwood {\em et al.} 1996, Larwood \& Papaloizou 1997). We have also 
examined the structure of warped discs orbiting about Kerr black holes
(the `Bardeen--Petterson effect') in a companion paper
to this one (Nelson \& Papaloizou 1999).

\end{document}